\documentclass[preprint,12pt]{aastex}

\newcommand{\SOHO}{{\it SOHO}}
\newcommand{\TRACE}{{\it TRACE}}
\newcommand{\STEREO}{{\it STEREO}}

\begin{document}
\baselineskip=20pt
\title{Observational Test of Coronal Magnetic Field Models \\
       I. Comparison with Potential Field Model}

\author{Yu Liu \& Haosheng Lin}
\affil{Institute for Astronomy, University of Hawaii \\
       34 Ohia Ku Street, Pukalani, HI 96768, USA}

\begin{abstract}
Recent advances have made it possible to obtain two-dimensional 
line-of-sight magnetic field maps of the solar corona from 
spectropolarimetric observations of the \ion{Fe}{13} 1075 nm forbidden 
coronal emission line. Together with the linear polarization measurements 
that map the azimuthal direction of the coronal magnetic field projected 
in the plane of the sky containing Sun center, these coronal vector 
magnetograms allow for direct and quantitative observational testing of 
theoretical coronal magnetic field models. This paper presents 
a study testing the validity of potential-field coronal magnetic field 
models. We constructed a theoretical coronal magnetic field model of 
active region AR 10582 observed by the SOLARC coronagraph in 2004 by using 
a global potential field extrapolation of the synoptic map of Carrington 
Rotation 2014. Synthesized linear and circular polarization maps from 
thin layers of the coronal magnetic field model above the active 
region along the line of sight are compared with the observed maps. 
We found that the observed linear and circular polarization signals 
are consistent with the synthesized ones from layers located just above 
the sunspot of AR 10582 near the plane of the sky containing the Sun 
center. 

\end{abstract}


\section{Introduction}
\label{sec:intro}

Understanding the static and dynamic properties of the solar corona 
is one of the great challenges of modern solar physics. Magnetic fields
are believed to play a dominant role in shaping the solar corona. 
Current theories also attribute reorganization of the coronal magnetic 
field and the release of magnetic energy in the process as the primary 
mechanism that drives energetic solar events. However, direct measurement 
of the coronal magnetic field is a very difficult observational problem.
Early experiments have demonstrated the feasibility of the measurement 
of the orientation of the coronal magnetic fields by observation of the 
linear polarization of forbidden coronal emission lines (CELs) in the 
visible and at IR wavelengths \citep{eddy1967,mickey1973,arnaud1982,
querfeld1982,tomczyk_et_al_2007}. Radio observations have also been 
successful in measuring the strength of the coronal magnetic field near 
the base of the solar corona \citep[e.g.,] [and references therein] 
{brosius_2006}. Direct measurement of the coronal magnetic field strength 
at a higher height by IR spectropolarimetry of the CELs was achieved 
only recently \citep{lin_et_al_2000, lkc_2004}. 
Without direct measurements, past studies involving coronal magnetic fields 
have relied on indirect modeling techniques to infer the coronal 
magnetic field configurations, including coronal intensity images observed in 
the EUV and X-ray wavelength ranges and numerical methods that reconstruct 
the three-dimensional coronal magnetic field structure by extrapolation 
and MHD (magnetohydrodynamics) simulation based on photospheric magnetic 
field measurements. Since experimental verification of theories and models 
is one of the cornerstones of modern science, the lack of observational 
verification of these indirect magnetic field inference methods that 
are in widespread use is a very unsatisfactory deficiency in our field. 

Our 2004 observations were obtained above active region AR 10582 right 
before its west limb transit. We have obtained the first measurement 
of the height dependence of the strength of the line-of-sight (LOS) 
component of the coronal magnetic field in this data, and it showed an 
intriguing reversal in the direction of the LOS magnetic field at a 
height of approximately $0.15\ R_{\odot}$ above the solar limb, as shown 
in Figures 4 and 5 of \cite{lkc_2004}. This feature and the observed 
linear polarization map have presented us with our first opportunity to 
carry out a comprehensive observational test of our coronal magnetic 
modeling methods in which the strength and direction of the  magnetic 
fields predicted by the models can be directly checked by the observations. 

Force-free extrapolation of photospheric magnetic fields is currently 
the primary tool for the modeling of coronal magnetic fields. However, 
it is not without limitations or uncertainties. For example, 
the force-free assumption does not hold true in the photosphere and low 
chromosphere, and possibly in the high corona above 2 $R_{\odot}$ 
\citep{gary_2001}. Moreover, a different assumption (current-free, linear and 
nonlinear force free) about the state of the electric current in the corona 
can lead to substantially different extrapolation results. Without direct 
magnetic field measurements, many fundamental questions concerning the 
basic assumptions and the validity of our tools cannot be addressed 
directly. To date, questions like ``Is a potential field approximation 
generally an acceptable approximation for coronal magnetic fields?" or 
``Do linear or nonlinear force-free extrapolations provide a more accurate 
description of the coronal magnetic field?" can only be addressed by 
visual comparison between the morphology of selected field lines of the 
extrapolated magnetic field model and observational tracers of coronal 
magnetic fields such as the loops seen in EUV images. However, these 
visual tests are qualitative and subjective. Furthermore, they assume 
the coalignment between the magnetic field lines and the loops in EUV 
images, which has not been verified observationally. 

As our first test, we attempted to address the question ``Is the potential 
field extrapolation generally an acceptable approximation for the coronal 
magnetic field?" This test was conducted by comparing the observed polarization 
maps of AR 10582 to those derived from a coronal magnetic field model 
constructed from the potential field extrapolation of photospheric magnetic 
field data. However, before we present our study, we should point out 
that because of the nature of our modeling tools and observational data, 
the results and conclusions of this research are subject to certain 
limitations and uncertainties. One of the intrinsic limitations of the 
observational data used in this study is that because of our single sight 
line from Earth to the Sun, the photospheric and coronal magnetic field 
observations cannot be obtained simultaneously. In the case of global 
coronal magnetic field models, the whole-Sun photospheric magnetic field 
data used as the boundary condition of the extrapolation can be 
obtained only over an extended period of time. Although the large-scale 
magnetic structure of nonflaring active regions may appear stable over 
a long period of time, high-resolution EUV observations have shown that 
the small-scale coronal structures are constantly changing. Thus, 
studies such as ours that compare coronal magnetic field observations 
and models constructed from photospheric magnetic fields inevitably 
are subject to uncertainties due to the evolution of the small-scale 
structures in the regions, and we should not expect a precise match 
between the observed and synthesized polarization maps. 

Another deficiency due to the single sight line of our observations
is the lack of knowledge of the source regions of the coronal radiation. 
This is perhaps the most limiting deficiency of the observations and 
models of this research. The uncertainty of the location of the source
regions associated with coronal intensity observations due to the low 
optical density of the coronal plasma and the resulting long integration 
path length is familiar. In the case of the coronal magnetic 
observations, the LOS integration problem prevents us from performing 
an inversion of the polarization data to reconstruct the three-dimensional 
magnetic field structure of the corona for direct comparison with those 
derived from extrapolations or MHD simulations.  On the other hand, since 
extrapolation techniques do not include the thermodynamic properties of 
the plasma in the construction of the coronal magnetic field models, 
they do not include information about the location of the 
CEL source regions either. Therefore, they cannot predict the 
intensity and polarization distribution of the CEL projected on the 
plane of the sky that are needed for direct comparison with the 
polarimetric observations. 

Without the information about the location of the source regions 
from the observational data and the extrapolated models, we adopted 
a trial-and-error approach in which synthesized linear and circular 
polarization maps were derived using empirical source functions
and the extrapolated potential magnetic field model, and were 
compared directly with those obtained from observations. Obviously, 
if acceptable agreement can be achieved with any of the models tested,
then we can argue with a certain degree of confidence that these models
are plausible models of the observed corona and that the potential 
field approximation is a reasonable approximation of the coronal 
magnetic fields. Nevertheless, we should emphasize that this 
trial-and-error approach is not an exhaustive search of all the 
possible source functions and therefore cannot provide a clear-cut 
true-or-false answer. In other words, even if no acceptable 
agreement can be found with all the model source functions we have 
considered, the potential field approximation still cannot be 
dismissed completely. 

\section{Data Analysis and Results}

\subsection{Modeling the Coronal Magnetic Fields}

\subsubsection{Evolutionary History of AR 10581 and AR 10582}

\begin{figure}[c] 
\epsscale{1.0}
\plotone{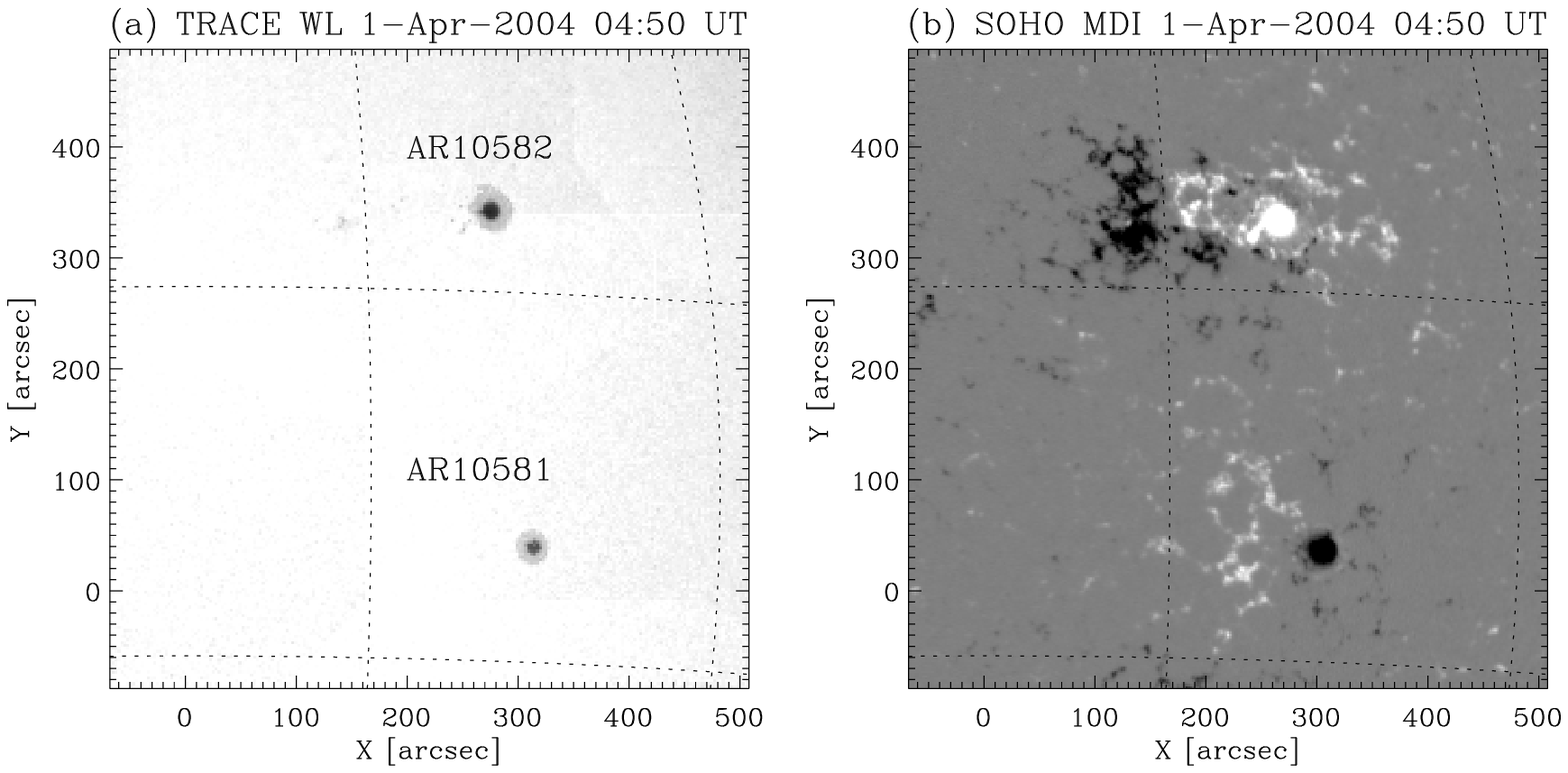}
\figcaption{\TRACE\ white-light image (left) and \SOHO/MDI 
magnetogram (right) of AR 10581 and AR 10582 observed on the solar 
disk show the regions' photospheric intensity and magnetic field 
configuration. 
\label{fig:context_info}
}
\end{figure}

\begin{figure}[c] 
\epsscale{0.75}
\plotone{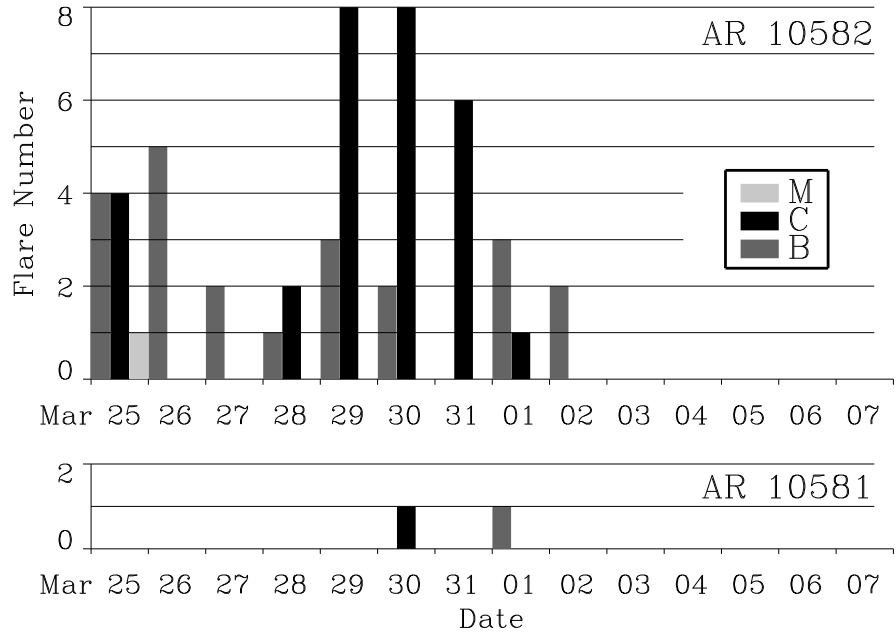}
\figcaption{Flare counts of AR 10581 and AR 10582 from 2004 March 25 to 
2004 April 07.
\label{fig:flare_counts}
}
\end{figure}

Although the purpose of this study is to test the validity of potential 
field approximation for AR 10582, examination of the photospheric magnetic 
field configuration and evolution of the regions should be informative 
and helpful for assessing the results of this study. Figure 
\ref{fig:context_info} shows the \TRACE\ (Transition Region 
and Coronal Explorer) white-light image and the \SOHO/MDI 
(Solar and Heliospheric Observatory/Michelson Doppler Imager) 
magnetogram of these regions on 2004 April 1. Their activity history 
during their disk transit is shown in Figure \ref{fig:flare_counts}. 
AR 10581 and 10582 first appeared on the east limb of the Sun on 2004 
March 23. Images taken by \SOHO/EIT (Extreme-ultraviolet Imaging Telescope) and \TRACE\ 
data showed that AR 10582 was initially very active and produced several 
C- and M-class flares between March 23 and April 1. AR 10581, on the 
other hand, produced only two small flares in the first week.  No flare 
activities were observed from either region in the four days before the 
SOLARC observation. Close examination of EIT and \TRACE\ data also showed 
that the large-scale configuration of these active regions did not change 
significantly during this period. 

\subsubsection{Extrapolated Potential Field Model of AR 10582}

\begin{figure}[c] 
\epsscale{1.0}
\plotone{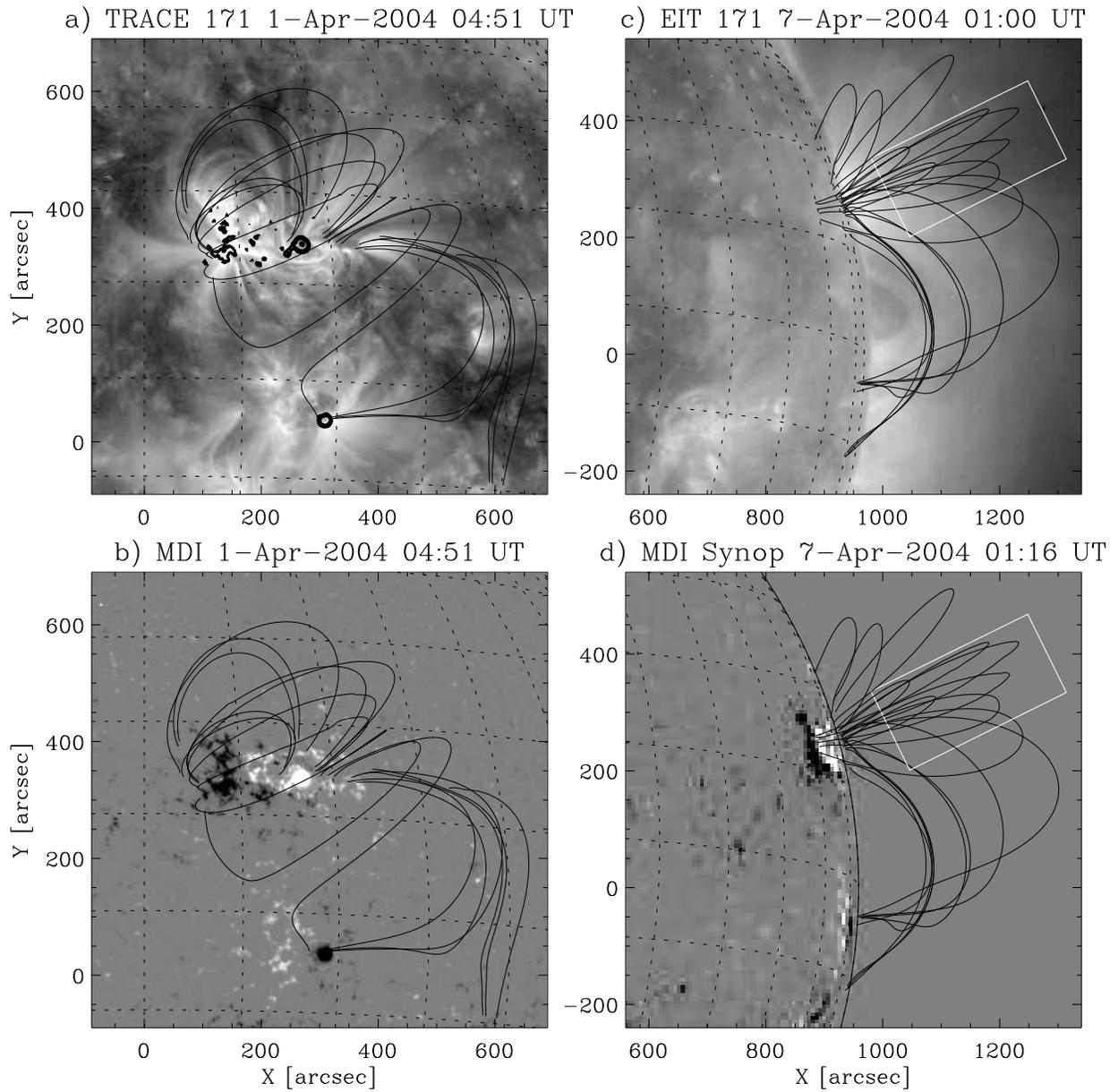}
\figcaption{Selected extrapolated magnetic field lines plotted over 
\ion{Fe}{9} 171 \AA\ images and \SOHO/MDI magnetograms of AR 10581 
and AR 10582 when these regions are observed on the solar disk (a and b)
and at the west limb (c and d).  
\label{fig:extrap_B}
}
\end{figure}

\begin{figure}[c] 
\epsscale{0.75}
\plotone{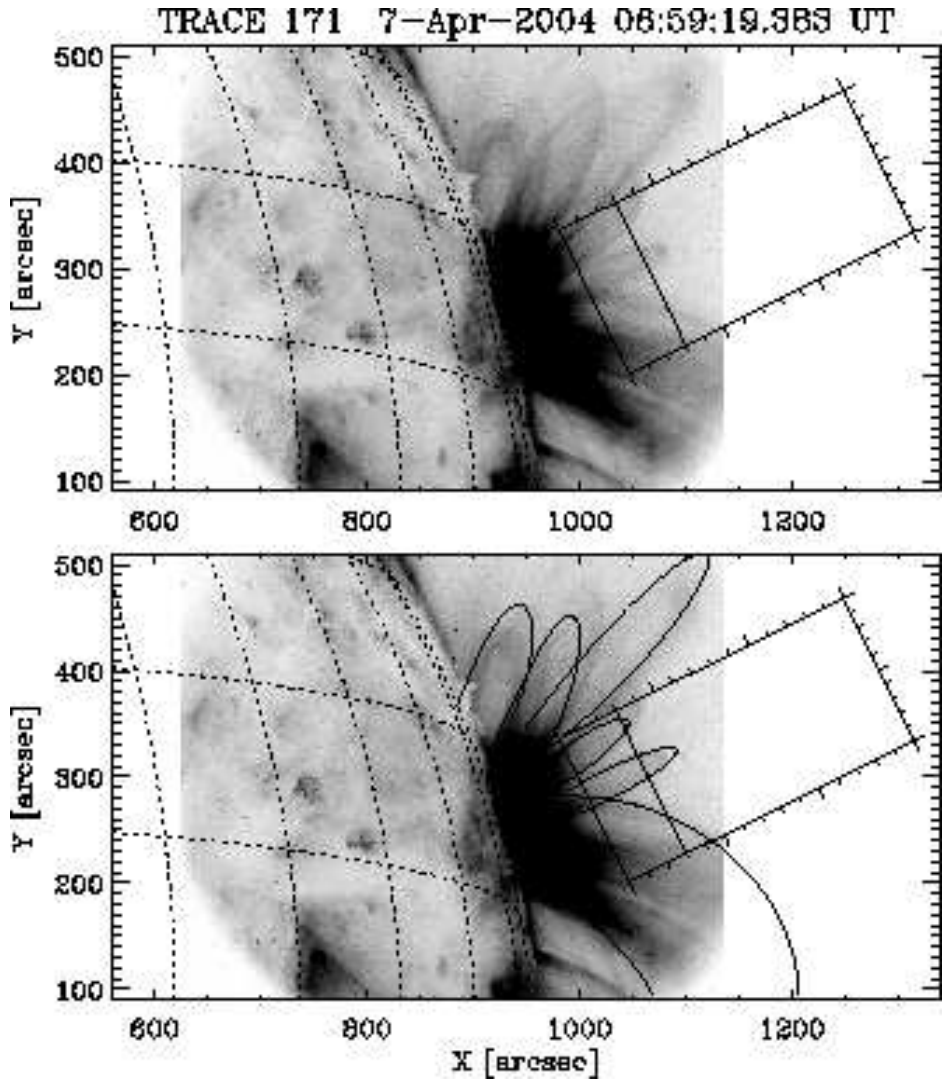}
\figcaption{top: \TRACE\ \ion{Fe}{9} 171 \AA\ image of AR 10582 about 3 
hr after the SOLARC magnetic field measurements were taken. The rectangle
marks the FOV of the SOLARC observations. The horizontal line within 
the SOLARC FOV indicates the location where a reversal in the direction 
of the longitudinal coronal magnetic field as a function of height above 
the solar limb was observed. bottom: Same as the top panel, with a subset 
of the magnetic field lines shown in Figure \ref{fig:extrap_B} plotted 
over the \TRACE\ image. 
\label{fig:trace_loops}
}
\end{figure}

We employed a global potential field extrapolation program based on 
the Green's Function method developed by \cite{schatten_et_al_1969} 
and \cite{sakurai_1982} for the construction of the coronal magnetic 
field \citep{liu_2002}. The magnetic synoptic map of Carrington Rotation 
2014 obtained by the \SOHO/MDI instrument was used as the boundary 
condition of the extrapolation. The first two panels, (a) and (b), of
Figure \ref{fig:extrap_B} show selected field lines from the 
extrapolated coronal magnetic field model of AR 10581 and AR 10582 
plotted over a \TRACE\ \ion{Fe}{9} 171 \AA\ image as they are viewed on 
the disk. Panels (c) and (d) show the same set of field lines overplotted on 
the \SOHO/EIT \ion{Fe}{9} 171 \AA\ images and \SOHO/MDI magnetograms 
when they transited the west limb. The field of view of the SOLARC LOS 
magnetogram observation is marked by the rectangular box in the figure. 
We found a general agreement between the orientation and distribution 
of the discernible EIT intensity loops during the time of the SOLARC 
observations, and the selected magnetic field lines from the extrapolated 
model are evident. The extrapolated magnetic field lines in AR 10582 are 
predominantly aligned along the east-west direction, while a set of field 
lines running along the north-south direction connecting the two active 
regions also coincide with a large north-south loop in the EIT images.  
Figure \ref{fig:trace_loops} shows a \TRACE\ high-resolution 
\ion{Fe}{9} 171 \AA\ image of AR 10582 taken about 3 hr after the 
limb spectropolarimetric observations had ended. We found that a 
subset of the extrapolated field lines appear to closely resemble 
the TRACE loops.  While the similarities in the morphology of the EUV 
intensity images and the extrapolated magnetic field lines seem to 
suggest that the extrapolated coronal magnetic field model is a fair 
representation of the magnetic field configuration of AR 10581 
and AR 10582, it is a subjective interpretation. 

\subsection{Observational Test of the Potential-Field 
            Coronal Magnetic Field Model}

\subsubsection{Synthesis of the Coronal Polarization Maps}

We have developed a program to calculate the LOS integrated linear 
and circular polarization signals at any point in the plane of the sky 
(POS) given the extrapolated coronal magnetic field model and the density 
and temperature distribution of the solar corona based on the classical 
theory of \cite{lin_casini_2000} for the forbidden CEL polarization. 
The formulae for the emergent Stokes parameter of the CEL are identical 
to those derived from a full-quantum mechanical formulation 
\citep{casini_judge_1999}, up to a proportional constant. However, 
as this classical formulation does not consider the effect of collisional 
depolarization, it may overestimate the degree of linear polarization at 
a lower height. The azimuth angle of the linear polarization predicted 
by our program should also be affected slightly when we integrate over 
a long path length, since the collisional depolarization effect may change 
the relative contribution of the sources along the LOS. 
\cite{judge_casini_2001} have also developed a CEL polarization synthesis 
program that includes the effect of collisions. We have implemented both 
programs to generate the linear and circular polarization signals from 
the extrapolated potential field model. 
We found that the synthesized polarization maps derived from these 
two programs are very similar, and analysis based on these two programs 
yielded the same results. The comparison of the synthesized polarization 
maps from these two programs will be presented when we present the 
study comparing the observed and synthesized polarization maps.

\subsubsection{Linear Polarization Maps}

The 2004 April 6 data include a linear polarization scan encompassing 
both AR 10581 and AR 10582. However, due to the long 
integration time required, only one circular polarization measurement 
was obtained above AR 10582.  Therefore, this study concentrates on 
the $320 \arcsec \times 160 \arcsec$ field with both circular and linear 
polarization measurements as marked by the rectangular area in Figures 
\ref{fig:extrap_B} and \ref{fig:trace_loops}. As it was mentioned in 
\S\ref{sec:intro}, the lack of knowledge of the coronal density and 
temperature distribution is the greatest uncertainty in our study. 
Nevertheless, based on decades of observations, we now know that strong 
coronal emissions are always associated with active regions. Furthermore, 
due to the small density scale height of the high-temperature emission 
lines, the contribution function of the CELs along the LOS is heavily 
weighted toward layers close to the POS containing Sun center. 
Therefore, we can expect that for isolated active regions, the 
forbidden coronal emission originates from a localized region near 
the POS containing Sun center during the active region's limb transit.  
According to the \SOHO/MDI white-light archive, there were no other 
active regions present on the solar disk on 2004 April 1, when AR 10581 
and AR 10582 were located approximately at disk center. Additionally, 
there was no evidence of new active regions emerging in the vicinity 
of these regions up to the day of the coronal polarization 
measurement.  Therefore, it is reasonable to assume that the polarized 
radiations we measured originated in the corona above these active 
regions near the POS containing Sun center. This expectation prompted 
us to test if an empirical source function constructed from a simple 
gravitationally stratified atmosphere and local magnetic field properties, 
such as the strength or the magnetic energy, can reproduce (if only 
qualitatively) the observed polarization maps.  

\begin{figure*}[c] 
\epsscale{1.0}
\plotone{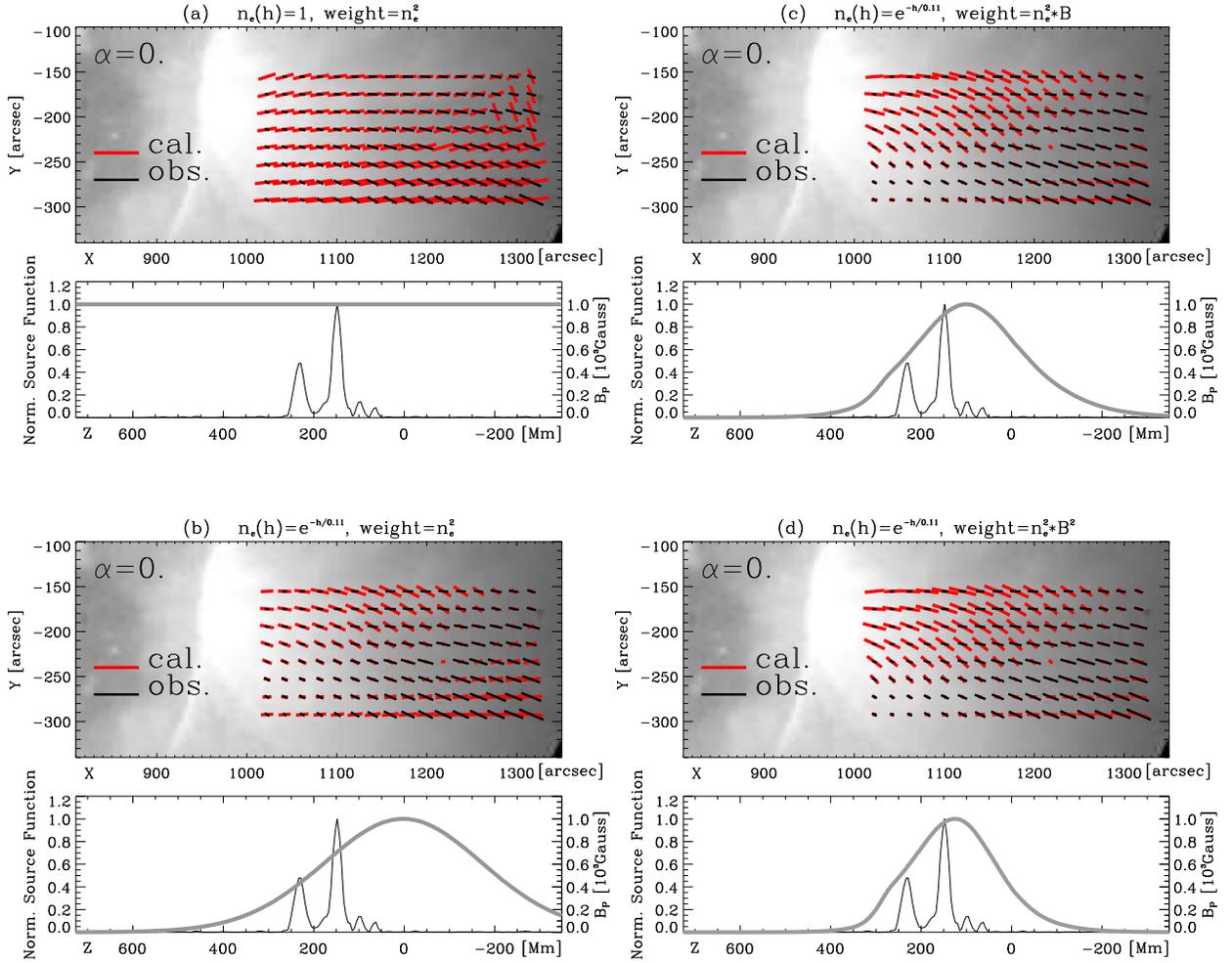}
\figcaption{Comparison between observed and synthesized linear polarization
maps with four empirical source functions: (a) reference source function with 
uniform density and temperature distribution, (b) source function based on 
a gravitationally stratified density distribution and uniform temperature, 
(c) gravitationally stratified density source function in (b) weighted by 
the local magnetic field strength, and (d) gravitationally stratified density 
source function weighted by the local magnetic field energy density, $B^2$. 
The observed (black lines) and synthesized (red lines) linear polarization 
maps are shown in the top figure in each panel, and the source functions 
(thick grey lines) from the center of the observed field along the LOS are 
shown in the bottom figure in each panel. The thin black lines in the bottom
figures show the strength of the observed photospheric magnetic flux. The 
source function in (a) has an equal contribution from every point in 
space. It is not a physically realistic model and is shown to demonstrate 
how a simple gravitationally stratified density distribution can affect 
the outcome of the simulation.
\label{fig:empirical_models}
}
\end{figure*}

For our first test, we constructed a linear polarization map using a 
source function ${W_{n_e}}(\bf{r})$ that is proportional to the square of the density of 
a spherically symmetric, gravitationally stratified density distribution 
with a density scale height of $h_0=83$ Mm 
\citep[$0.11\ R_{\odot}$,][]{lang_1980}.  That is, 
\begin{eqnarray}
\label{eq:source_function_1}
W_{n_e}({\bf r}) = e^{-2h({\bf r})/h_0},
\end{eqnarray}
where ${\bf r}$ is the three-dimensional position vector in the heliocentric 
coordinate system and $h({\bf r})$ is the height at ${\bf r}$. 
The temperature of the corona was assumed to be constant and did not affect 
the source function. The magnetic properties of the corona are not included 
in this model either. This is similar to the atmospheric model used in the 
polarization synthesis performed by \cite{judge_et_al_2006}. The resulting 
normalized source function along the LOS in the center of the SOLARC field 
of view and the synthesized linear polarization map are shown in panel 
(b) of Figure \ref{fig:empirical_models}.  Note that because 
AR 10582 was located between W70 to W80  longitude, this density-only source 
function has its maximum located outside of the active region. Although 
$W_{n_e}({\bf r})$ is obviously a gross simplification of the magnetic coronal 
atmosphere, and we should not expect to see good agreement between the observed 
and synthesized linear polarization maps based on $W_{n_e}({\bf r})$ alone 
(as demonstrated in the top figure in panel (b) of Figure 
\ref{fig:empirical_models}),  this test is a necessary step in our systematic 
trial-and-error study. Furthermore, 
the importance of its inclusion in the estimate of the source function 
can be seen when we compare the linear polarization maps constructed with 
and without it (top figure in panel (a) of Figure \ref{fig:empirical_models}). 

In our next test, we examined if there is any simple relationship 
between the CEL source function and the magnetic properties of the corona.  
Because of the observed correlation between the strong  CEL radiation and 
active regions, it is only logical to test if a source function with its 
amplitude proportional to the local magnetic field strength or magnetic 
energy can reproduce the observed polarization signals. To test this idea,
we multiplied the source function that was constructed from the uniform 
temperature, gravitationally stratified atmosphere model by the local 
magnetic field strength $B$. Another model used the local magnetic field 
energy density $B^2$ as the additional magnetic weighting function. 
Accordingly, the two magnetic source functions are expressed by
\begin{eqnarray}
W_{B}({\bf r}) = e^{-2h({\bf r})/h_0} B
\end{eqnarray} 
and
\begin{eqnarray}
W_{B^2}({\bf r}) = e^{-2h({\bf r})/h_0} B^2.
\end{eqnarray} 
The additional magnetic constraints restrict the source function to a 
more localized region around the active region, and with a spatial 
scale comparable to that of the active region.  However, the linear 
polarization maps derived using these source functions still are not 
in good agreement with the observed one. These are demonstrated in 
panels (c) and (d) of Figure \ref{fig:empirical_models}. 

Although none of the linear polarization maps we have constructed 
so far can be considered to be in good agreement with the observation, 
it can be argued that the maps derived from the two source functions 
with magnetic constraints appear to better match the observed one, 
especially at the lower right-hand corner of the field, where the 
degree of linear polarization is the highest. Therefore, we suspect 
that the assumption about the magnetic dependence of the CEL source 
function is in general valid. However, the correlation may be occurring 
at a spatial scale smaller than that of the active regions.  Because 
space EUV observations have shown that radiation from the emission-line 
corona originates from loop-like structures with a characteristic 
size much smaller than the characteristic size of the active regions, 
it should not be surprising that the linear polarization maps produced 
by the broad source functions do not agree well with the observation.
This reasoning prompted us to experiment with source functions with
a spatial scale approximately equal to a few times the characteristic 
width of the coronal loops to test if better agreement between the 
observed and synthesized polarization maps can be achieved.  

\begin{figure*}[c] 
\epsscale{1.0}
\plotone{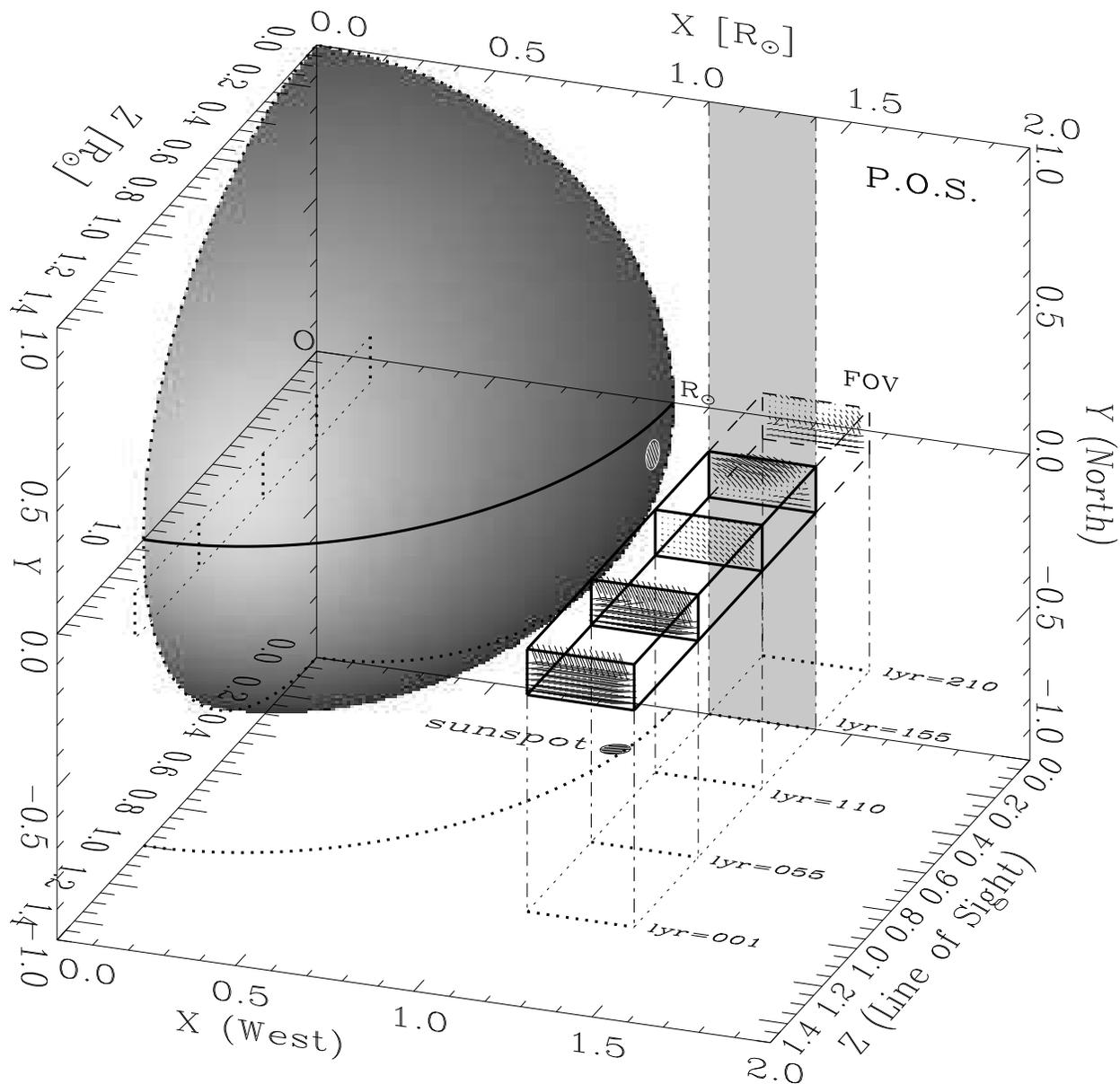}
\caption{
Three-dimensional plot to illustrate the locations of the thin layers along 
the LOS, and their relationship to the solar sphere and the sunspot 
of AR 10582. The Sun is represented by the shaded quarter 
sphere. The observing LOS is along the $Z$ axis. Layer 155 is located 
in the POS containing the Sun center. 
\label{fig:obs_geometry}
}
\end{figure*}

Since we do not have information about the location of the source regions 
of the observed \ion{Fe}{13} 10747 \AA\ line emission at this spatial 
resolution, we constructed the linear polarization map of 205 layers 
along the LOS within the SOLARC FOV for comparison with the observed 
linear polarization map. The separation between the synthesized layers 
is 4.5 Mm, the resolution of the potential field extrapolation calculation. 
The first and last layers are located 720 Mm (or about 1000$\arcsec$) in 
front and 250 Mm (or about 340$\arcsec$) behind the POS 
containing Sun center, respectively. The locations of these layers with 
respect to the solar sphere, the observed active region, and the LOS 
of the observer are illustrated in Figure \ref{fig:obs_geometry}.  
We used a one-dimensional Gaussian function with a full width at 
half-maximum (FWHM) in the LOS direction equal to a few times that 
of the characteristic coronal loop size to model the source function. 
For this test, we adopted the 8 Mm characteristic loop width of the 
EUV \ion{Fe}{14} 28.4 nm line derived by \cite{aschwanden_et_al_2000}. 
This line is chosen because its ionization temperature of $2.2\times10^6$ 
K is closer to that of the IR \ion{Fe}{13} 10747 \AA\ line 
($T_{\rm ion} = 1.7 \times 10^{6}$ K) than the other \SOHO/EIT EUV 
lines. We calculated and compared the polarization maps with the FWHM 
of the source function set from 1 to 14 times the characteristic loop 
width of the \ion{Fe}{14} 28.4 nm lines and found no significant 
difference between the polarization maps. This is expected, since the 
coronal magnetic field should vary slowly in space as it expands to 
fill the entire coronal volume. Finally, because of this lack of 
sensitivity in the synthesized polarization signals to the variations 
of the FWHM of the source function, we used a nominal 56 km FWHM 
source function (or 7 times the FWHM of the \ion{Fe}{13} 28.4 nm loops) 
for the calculation of the synthesized polarization maps in the rest 
of the paper. 

\begin{figure}[c] 
\epsscale{1.0}
\plotone{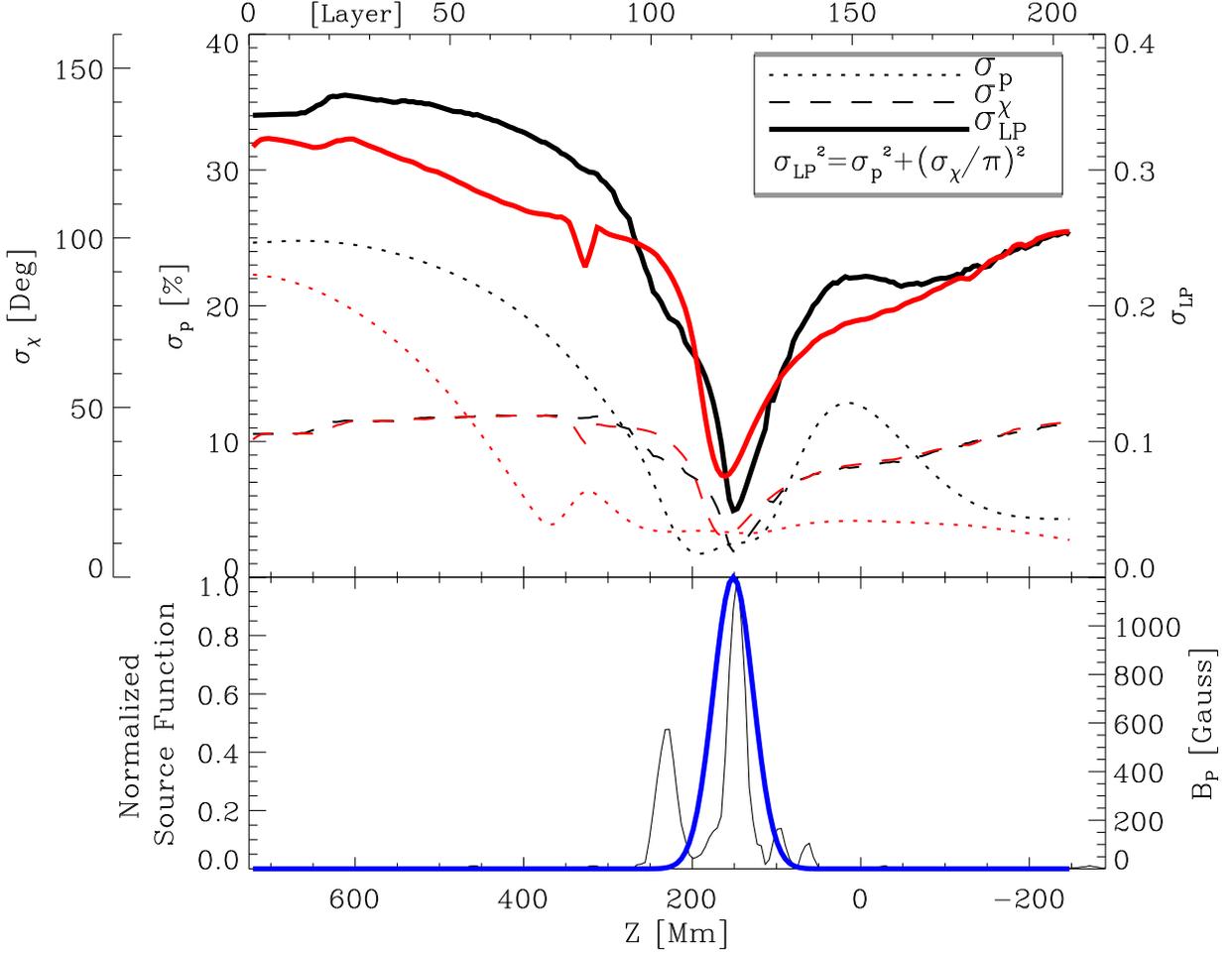}
\figcaption{
top: The rms errors between the synthesized and observed linear 
polarization amplitude, $\sigma_p$ (dotted lines), azimuthal angle, 
$\sigma_\chi$ (dashed lines), and the combination of the two, 
$\sigma_{LP}$ (thick solid lines). Results derived with the coronal 
polarization synthesis program developed by the authors and 
\cite{judge_casini_2001} are shown in black and red, respectively.
Best-fit position for linear polarization occurs at approximately 
layer 120, right above the sunspot. 
bottom: The thin solid line shows the magnitude of the LOS magnetic 
field of the photosphere. The thick blue line shows the source 
function along the LOS for layer 120. 
\label{fig:fit_errors}
}
\end{figure}

\begin{figure*}[c] 
\epsscale{1.0}
\plotone{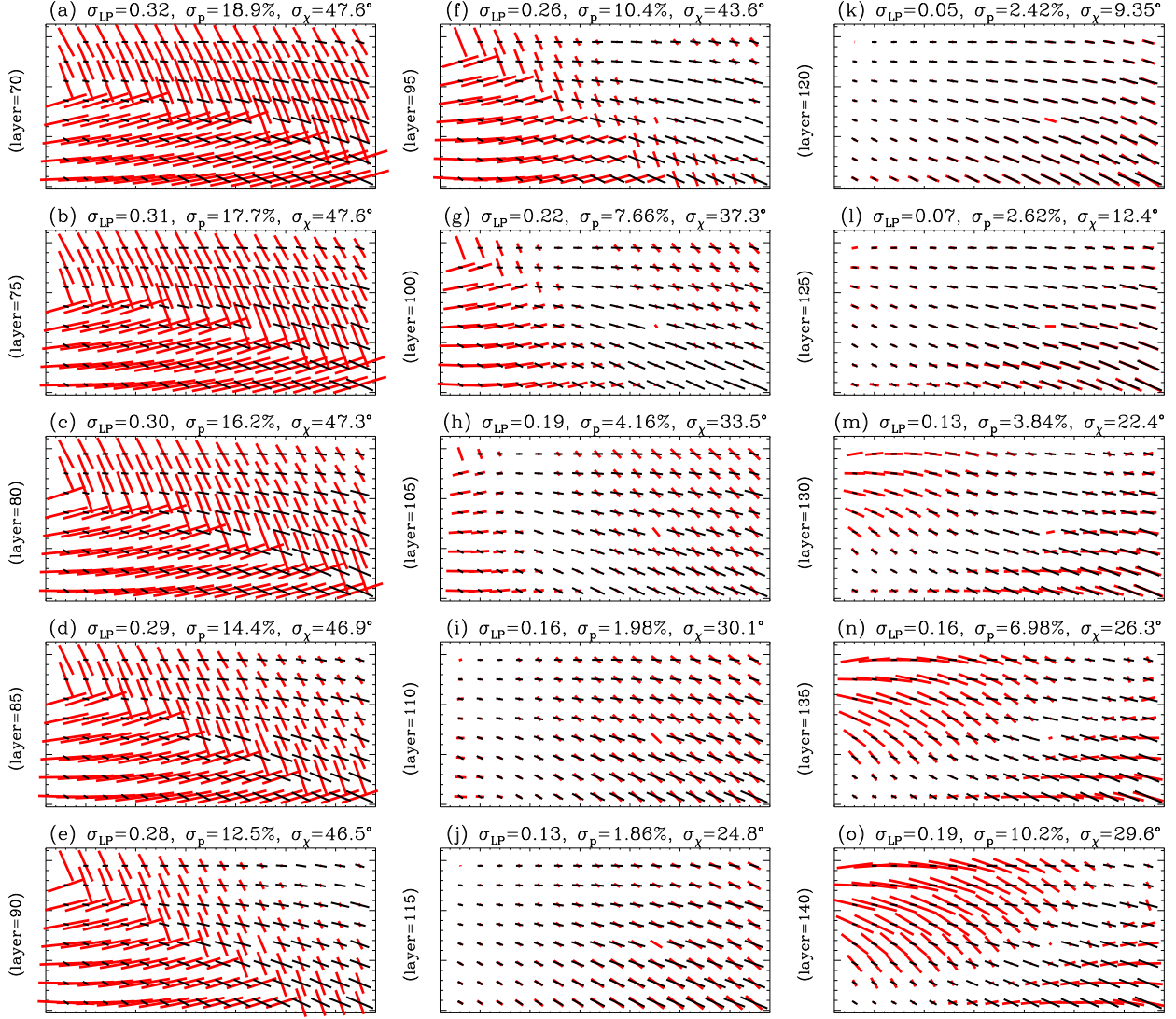}
\figcaption{
The observed (black lines) and synthesized (red lines) linear polarization
maps derived with our classical synthesis program from 15 layers along the 
LOS near the active region. Layer 120 is the layer with the smallest rms 
error. 
\label{fig:linear_polarization_maps}
}
\end{figure*}

We evaluated the quality of the fit between the synthesized and observed 
polarization maps from the rms difference between the degree of polarization 
$p$ and the azimuthal angles $\chi$ of the linear polarization direction 
projected in the POS, $\sigma_p$ and $\sigma_\chi$, respectively. 
Figure \ref{fig:fit_errors} shows $\sigma_p$ and $\sigma_\chi$ as functions 
of $z$ along the LOS. Results derived from our own collisionless classical 
formulation are shown by the black lines, and those derived with Judge \& 
Casini's code (2001; hereafter referred to as the JC synthesis code) are shown 
in red. Figure \ref{fig:fit_errors} shows that the minimum of 
$\sigma_p$ and $\sigma_\chi$ both 
occur near the layer right above the sunspot of the active region. Figure 
\ref{fig:linear_polarization_maps} shows 15 synthesized linear polarization 
maps calculated with our classical synthesis code from layer 70 to layer 140, 
with an interval of 5 layers, plotted over the observed map. The best fit, 
determined from the total rms error $\sigma^2_{lp} = \sigma^2_p + 
(\sigma_{\chi}/\pi)^2$ of the linear polarization maps, occurs around layer 
120, right above the sunspot.

\subsubsection{Comparison with Judge \& Casini's Coronal Polarization 
               Synthesis Code}

\begin{figure}[c] 
\epsscale{0.75}
\plotone{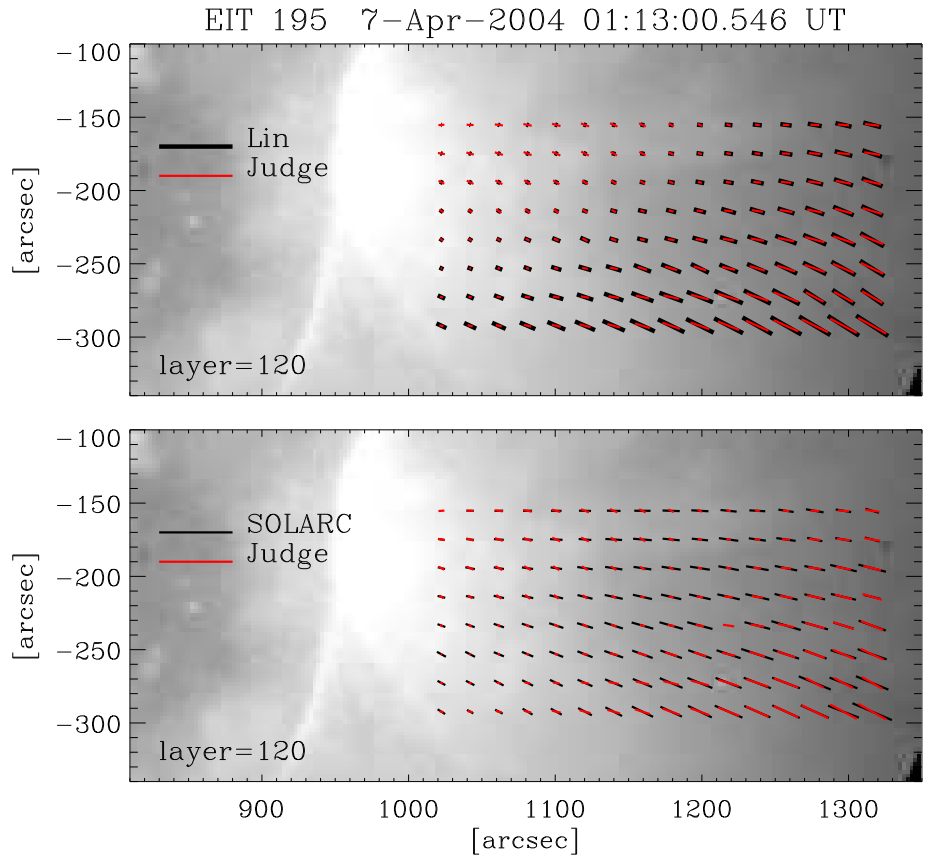}
\figcaption{
top: The linear polarization maps derived from Judge \& Casini's
synthesis program (red lines), and from Liu and Lin's classical 
synthesis program (black lines). bottom: Comparison of the linear 
polarization map derived from Judge \& Casini's program and that 
observed by SOLARC. 
\label{fig:judge_lp}
}
\end{figure}

Figure \ref{fig:judge_lp} shows the linear polarization map of layer 
120 derived with the JC synthesis code compared with that derived from 
our collisionless code, and with observations. Apparently, the JC synthesis 
code consistently predicts a smaller linear polarization amplitude 
compared with our classical theory, as expected, and produced better 
agreement with the observed linear polarization map at the lower part 
of the field. However, there are still significant disagreements in the 
upper part of the field, where both the JC synthesis code and our own program 
overestimated the degree of linear polarization. This may be due to the 
larger measurement errors associated with the small amplitudes of the 
observed linear polarization, and their significance should not be 
overstated. Close examination of the data shows that the larger rms 
error shown in Figure \ref{fig:fit_errors} in the JC results is due to 
larger errors in this region. Therefore, we do not consider that our 
collisionless synthesis actually provides better agreement with the 
observations. 

In view of the inherent uncertainties in the modeling process discussed 
in \S\ref{sec:intro}, we do not feel that a meaningful quantitative 
comparison between these two methods can be justified at this point. 
Nevertheless, Judge \& Casini's program with collisional depolarization 
is definitely a more complete description of the physical processes in 
the atmosphere of the solar corona and should be preferred. Future 
observations at a lower height, where the collisional depolarization effect is 
expected to be more important, should allow us to clearly distinguish 
between the effect of collisional depolarization and uncertainties in 
the modeling process. 

\subsubsection{Strength and Reversal of Line-of-Sight Magnetic Fields}

Our analysis so far has demonstrated that the potential field 
model can reproduce the observed linear polarization maps. 
Is this a coincidence? To answer this question, it is interesting 
to first note that the layers of best fit for both the degree and 
azimuthal angle of the linear polarization occur at approximately 
the same location near the region of the strongest photospheric 
magnetic fields. Since $\sigma_p$ and $\sigma_{\chi}$ were determined 
independently, the probability that these two parameters reach minimum 
at approximately the same location and at the region with the strongest 
photospheric magnetic flux purely by chance should be very low. 
Therefore, we believe that our analysis of the linear polarization 
maps support the idea that the potential field model is a good 
approximation of the real coronal magnetic field of AR 10582. 
Furthermore, if an agreement between the modeled and observed height 
dependence of the LOS component of the coronal magnetic field can be 
found, then the validity of the potential field model, at least 
as a first order approximation for stable active regions, can be 
strongly argued. 

\begin{figure*}[c] 
\epsscale{1.0}
\plotone{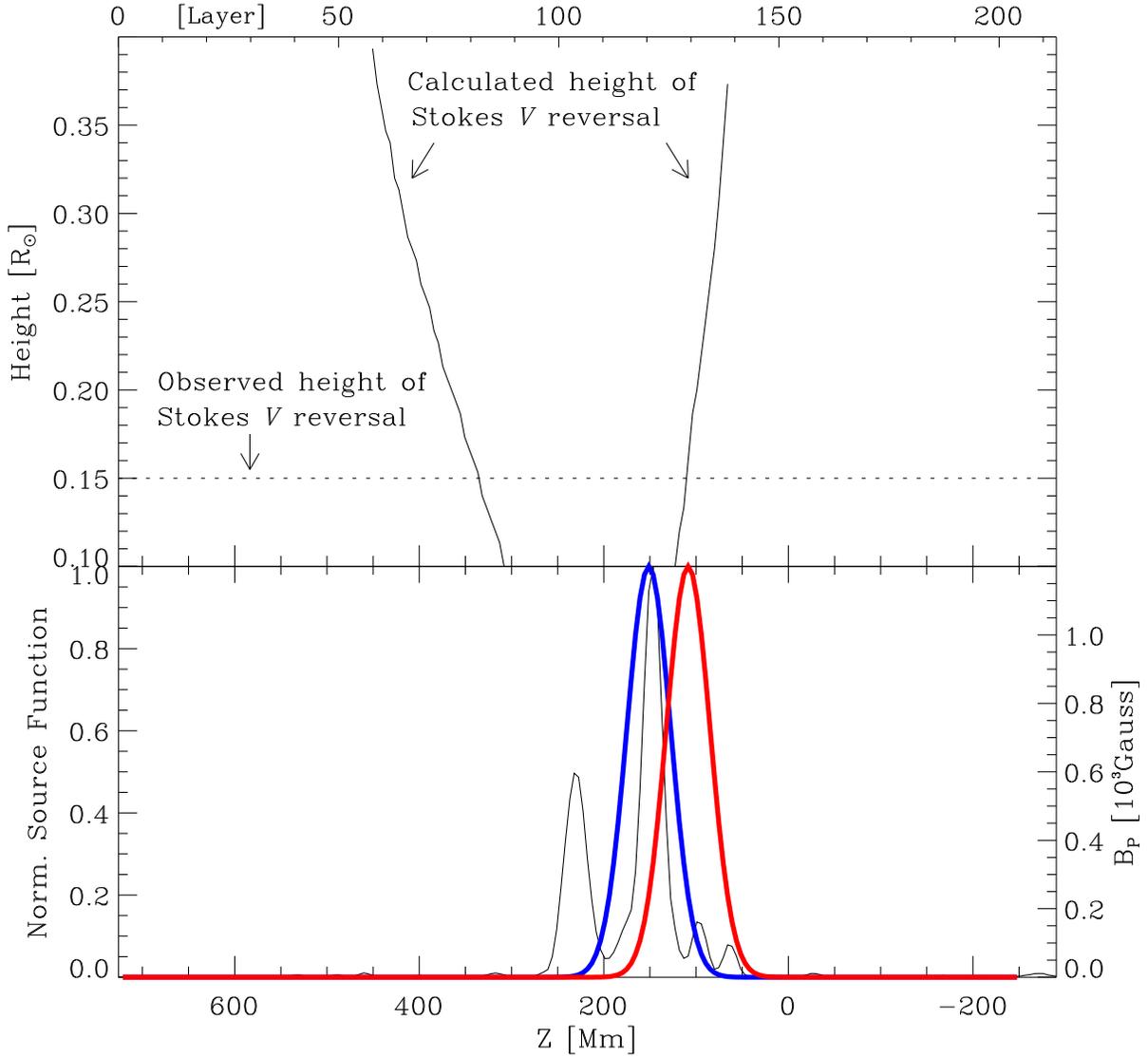}
\caption{
top: The calculated height of Stokes $V$ reversal as a function of height 
from the limb derived from the potential field model. 
bottom: Similar to Figure \ref{fig:fit_errors}, the thin solid line shows 
the magnitude of the LOS magnetic field of the photosphere, and the thick 
blue line shows the source function of layer 120 where the best fit for the 
linear polarization map occurs. The thick red 
line shows the source function along the LOS for layer 130. 
\label{fig:StokesV_Reversal}
}
\end{figure*}

\begin{figure}[c] 
\epsscale{1.0}
\plotone{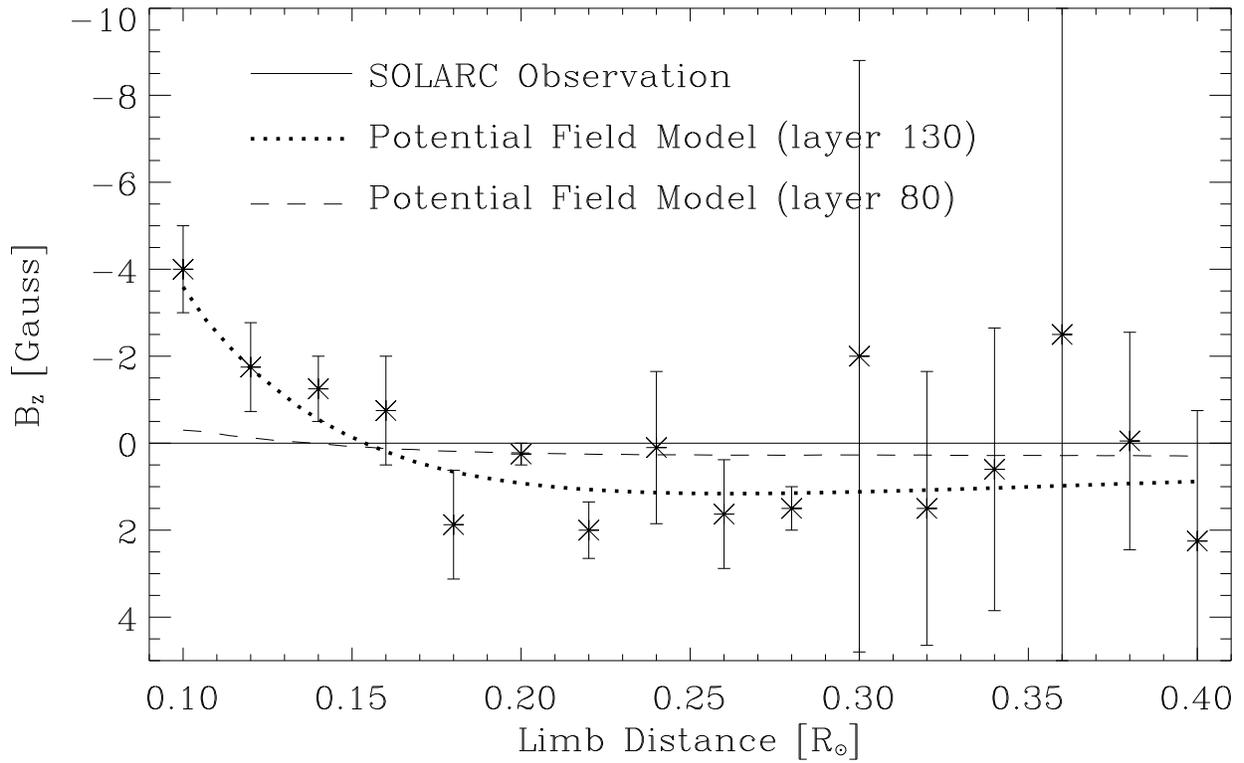}
\figcaption{
The observed (solid line with star and error bars) and synthesized
net circular polarization signals from the source layer 130 (dotted line) 
and layer 80 (dashed line) are plotted as a function of distance from 
the solar limb.  
\label{fig:Synthesized_StokesV}
}
\end{figure}

To check if the observed Stokes $V$ reversal can be reproduced, we 
derived the height of Stokes $V$ reversal $H_0$ for each of the 205 
56-Mm-FWHM layers along the sight line. The LOS components of the 
magnetic field $B_z$ in the central $320 \arcsec \times 80 \arcsec$ 
region of each layer were averaged in the north-south direction (or 
the tangential direction with respect to the local solar limb) to 
simulate the spatial averaging performed on the observation data. 
The result is  shown in Figure \ref{fig:StokesV_Reversal}. 
Since the dominant magnetic structure around the active region in our 
FOV consists of magnetic loops oriented along the east-west direction, 
there were two locations with $H_0 = 0.15\ R_{\odot}$, one due to the 
front (closer to the observer) portion of the loops near layer 80, and 
one located in the back of the loops, at layer 130. Note that layer 130 
is much closer to the sunspot of AR 10582 and the maximum of the empirical 
source functions $W_B$ and $W_{B^2}$ shown in Figure 
\ref{fig:empirical_models}. In comparison, the amplitudes of 
$W_{n_e}$, $W_B$, and $W_{B^2}$ at layer 80 are only about 0.15, 0.1, 
and 0.05, respectively (Figure \ref{fig:empirical_models}). 
We calculated the net Stokes $V$ signals as a function of height above 
the limb for layer 130 and found that they agree well with the observed 
signals as shown in Figure \ref{fig:Synthesized_StokesV}. 
On the other hand, this is not the case for layer 80.
We conducted the same analysis using the JC synthesis code and 
produced virtually identical results. 
Therefore, it is more likely that the dominant source of the Stokes 
$V$ signals originates from around layer 130. 

The blue and red Gaussian curves in Figure \ref{fig:StokesV_Reversal} 
mark the locations where the best fits to the 
linear and circular polarization observation occur, respectively.  
As is clearly shown in this figure, layer 130 is only about 50 Mm 
away from the location with the best fit of the linear polarization 
maps. Given the proximity of the locations of the best fit of the three
parameters ($p$, $\chi$, and $H_0$) and the strongest magnetic feature
of the active region, we concluded that this comparative study demonstrates
the validity of potential field extrapolation as a tool for the modeling 
of the coronal magnetic field. 

\section{Summary, Discussions, and Conclusions}

This research examines observationally the validity of current-free,
force-free potential-field approximation for coronal magnetic fields. 
We conducted a study comparing observed and synthesized spatial 
variations of the linear and circular polarization maps in the corona 
above active region AR 10582, which after a week of extensive flaring 
activities should have settled into a minimum energy configuration that 
could be adequately modeled by a potential field model. The coronal 
magnetic field model used for this study was constructed from a global 
potential field extrapolation of the synoptic photospheric magnetogram 
of Carrington cycle 2014 obtained by the \SOHO/MDI instrument. 
Because the most important source of error of this type of study is the 
uncertainty of the location of the source function of the coronal 
radiation, we first tested three analytical, but empirically determined, 
source functions.  These simple source functions are based on a 
gravitationally stratified atmospheric density model with a uniform 
temperature in the entire modeled volume, supplemented by magnetic 
weighting functions based on the observational impression that, at 
least at the length scale of the typical active region size, CEL 
radiation seems to be correlated with the strength of the photospheric 
magnetic fields. We found that none of these empirical source functions 
can adequately reproduce the observed linear polarization maps, although 
it seems that the source functions that include both density and magnetic 
fields produced slightly better results. Based again on the observational 
impression that the coronal intensity structures have a spatial scale much 
smaller than that of the active regions, we then compared the observed 
polarization maps with those constructed from thin (56 Mm FWHM) layers along 
the LOS. In this analysis, we found that polarization maps originating 
from layers located near the sunspot of the region are in reasonable
agreement with the observed ones. However, the best fit for linear and
circular polarization did not occur at the same layer. They are separated 
by a distance of about 50 Mm. 

Does the small discrepancy between the best-fit locations of the linear 
and circular polarization weaken the support for the potential field 
extrapolation? As we have discussed in \S\ref{sec:intro}, many 
uncertainties conspire to limit the precision of this study. For example, 
the difference may be due to the evolution of the small-scale photospheric 
magnetic field of the active region, and we do not have any observational 
information that we can use to test this possibility.  The assumption of 
uniform temperature distribution in our source function is certainly not 
a physically realistic assumption. Therefore, it is not possible to assess 
the significance of the small difference in the location of the source 
regions of the linear and circular polarization.  Furthermore, in addition 
to density and temperature, the source functions of CEL linear and circular 
polarization depend on different components of the coronal magnetic field. 
So it is in fact physically reasonable that we would find the linear and 
circular polarization signals originate from slightly different locations. 
Finally, because the three parameters ($\sigma_p$, $\sigma_\chi$, and $H_0$) 
we used to evaluate the quality of the fit were obtained independently, 
the statistical significance that all three parameters reached minimum 
at approximately the same location near the strongest photospheric magnetic 
feature of the active region cannot be dismissed as pure coincidence. These 
considerations lead us to conclude that potential field extrapolation can 
be used to provide a zero-order approximation of the real solar corona 
if the active region is in a relatively simple and stable configuration.
Additionally, this study suggests that, at least for isolated active
regions, CEL radiation may originate from a region close to the strongest 
photospheric magnetic feature in the active region with a small spatial 
scale comparable to the characteristic size of the coronal loops seen in 
the intensity images. If this is confirmed, then a single-source inversion
to infer the magnetic field directly from the polarimetric observation such
as that proposed by \cite{judge_2007} may be justified.  

Our conclusion about the viability of potential field extrapolation 
as a coronal magnetic field modeling tool for stable active regions 
is supported by a study by \cite{riley_et_al_2006}, in which the 
coronal magnetic field configuration derived from a potential field 
model was found to closely match that derived from a MHD simulation 
in the case of untwisted fields. Nevertheless, we should emphasize 
that our conclusion is derived from a single observation of a simple and 
stable active region. Clearly, more observations and model comparison are 
needed for a more comprehensive test of this result. 

Can the potential field approximation be used to model more complicated 
active regions? Using radio observations, \cite{lee_et_al_1999} found 
that a force-free-field model yields better agreement between the 
temperatures of two isogauss surfaces connected by the modeled field 
lines of an active region with strong magnetic shear. This study thus
provides observational evidence against the use of potential field 
approximations for the modeling of complex active regions. Therefore,
linear and nonlinear force-free extrapolations should be employed in 
future testing of theoretical coronal magnetic field models using the 
IR spectropolarimetric observations to study if these models can offer 
a better description of the observed coronal fields. 

Can we distinguish the potential coronal magnetic field configurations 
from the non-potential ones with the spectropolarimetric observations 
of the coronal emission lines? In a numerical study, \cite{judge_2007} 
has demonstrated the sensitivity of LOS-integrated 
coronal polarization measurements to the electric current in the corona  
using theoretical coronal magnetic field models as input.  Therefore, we 
should expect to find better agreement between observed and synthesized 
polarization maps for more complex active regions with linear or nonlinear 
force-free magnetic field models. Work to model AR10582 using the force-free 
extrapolation method is already underway, and we should be able to address 
this question in the near future. Since all extrapolation methods are 
subject to the ambiguities problem of the source regions, we will also 
employ MHD simulations that include both the magnetic and thermodynamic 
properties of the corona in the calculation to help resolve this problem. 
These are research activities that we will be pursuing in the near future 
as the solar cycle evolves toward the next solar maximum and more coronal 
magnetic field data become available. 

The greatest difficulty of this study is the uncertainty of the location 
of the source function due to the long integration path along the LOS.
However, this is not a difficulty affecting only the interpretation of 
coronal magnetic field measurements. It affects the intensity observation 
as well, and is the primary reason that years into the operation of 
\SOHO/EIT and \TRACE, we still cannot deduce 3-D intensity and temperature 
structure of the corona using data from these instruments. Fortunately, 
this deficiency in our observing capability may finally be removed with 
the recent launch of the \STEREO\ mission (Solar TErrestrial RElations 
Observatory).  For the resolution of the LOS 
integration problem in polarimetric observations, \cite{kramar_et_al_2006} 
have demonstrated the promising potential of vector tomography techniques.  
While stereoscopic coronal magnetic field observations will not be 
realized any time soon, this method can be applied to observations
obtained over periods of several days during the limb transit of active 
regions, provided that the active regions are in a stable condition. 
This is perhaps the best observational tool available for the resolution 
of the LOS integration problem in the near future. 

\acknowledgements
The authors would like to thank Phil Judge for generously providing 
the polarization synthesis codes and for helping to implement them. 
Y. L. thanks Tongjiang Wang for many discussions on this work.
The authors also gratefully acknowledge helpful comments from the 
anonymous referee that greatly improved the presentation of the paper.
EIT data is courtesy of the \SOHO/EIT consortium. \SOHO\ is a 
project of international cooperation between ESA and NASA. This 
research is funded by NSF ATM-0421582 and NASA NNG06GE13G.


{}

\clearpage


\clearpage

\clearpage

\clearpage

\clearpage

\clearpage

\clearpage

\clearpage

\clearpage

\clearpage

\clearpage

\clearpage

\end{document}